\documentclass{elsart3}
\usepackage{epsfig}
\usepackage{sidecap}

\begin{document}

\begin{frontmatter}

\title{Embedded metal cluster in strong laser fields}
\author{F. Fehrer$^1$, P. M. Dinh$^2$, P.-G. Reinhard$^{1,2}$, and
        E. Suraud$^{1,2}$}
\address {$^1$Institut f\"ur Theoretische Physik, Universit\"at Erlangen
              Staudtstr.\ 7, D-91058 Erlangen, Germany\\
        $2$Laboratoire de Physique Th\'eorique,  UMR 5152,
        Universit{\'e} Paul Sabatier,
        118 Route de Narbonne, F-31062 Toulouse, cedex 4,
        France}
\begin{abstract}
We discuss microscopic mechanisms of the violent dynamics following
strong laser excitation of a metal cluster embedded in a rare gas
matrix, taking as test case Na$_8$@Ar$_{434}$. This covers at least
two aspects:
first, it represents the typical experimental situation of metal
clusters handled in raregas matrices or a finite drop of surrounding
raregas material,
and second, it serves as a generic test case for highly excited
chromophores in inert surroundings addressing questions of energy
transport and perturbation of the medium.
We simulate the process up to 10 ps using a mixed quantum mechanical
(for the electrons) and classical (ions and atoms) approach and
analyze the emerging dynamics with respect to all basic constituents:
cluster electrons, cluster ions, and matrix atoms.  We find several
stages of relaxation taking place with time scales from a few fs to
over a few ps and much slower processes remaining for long after the
simulation. A particularly interesting aspect is that the surrounding
raregas material stabilizes a highly charged metal cluster which would
otherwise explode without delay.
\end{abstract}

\end{frontmatter}


\section{Introduction}

Nanotechnology, which was predicted long ago to stimulate an
industrial revolution \cite{Fey60}, now constitutes a fast developing
research field, for example to increase the control of material at
nanometer scale \cite{Ebe02}. Especially metal clusters have attracted
much attention in the past decades
\cite{Kre93,Hee93,Bra93,Hab94ab,Eka99,Rei03a}. Among their many
interesting characteristics and applications, their optical properties
are particularly appealing \cite{Kre93}. These are dominated by the
Mie surface plasmon which plays a key role in all dynamical regimes
\cite{Rei03a}. It provides a strong coupling to photons in a very
narrow frequency window which makes clusters an ideal laboratory for
laser induced non-linear dynamics \cite{Cal00,Teu01}. The plasmon also
serves as a versatile handle for pump and probe experiments
\cite{Bes99,And02,Doe05b,Din05,And06}, and it provides the key
mechanism in driving the hefty Coulomb explosion of clusters
\cite{Dit96,Buz96}.  For a recent summary on this great variety of
phenomena of the dynamics of free metal clusters, see
\cite{Rei04r,Bel04r}.
An even richer variety of phenomena emerges for metal clusters in
contact with other materials, either embedded in a matrix or deposited
on a surface. There is, e.g., second harmonic generation
\cite{Goe95,Bal00,Koh00}, dedicated shaping of clusters with intense
laser pulses \cite{Sei00,Oua05a}, and many more new phenomena emerging
from the combination, for a review see \cite{Bin01}.  Thus the proper
combination of materials is a key issue in nano-technology
\cite{Mil99}.  There are also practical reasons for dealing with such
compounds. The substrate simplifies the handling of the clusters such
that many interesting experiments can only be done with clusters in
contact with a carrier material \cite{Nil00,Leh00,Gau01,Die02}. Simple
compounds are also useful model systems for principle studies. For
example, clusters in contact with insulators can serve as a versatile
model system for chromophores which can be used as indicators in
biological tissues \cite{May01,Dub02} or for studies of radiation
damage in materials \cite{Bar02b,Niv00a}. The latter application deals
with high energy density related to intriguing questions of energy,
momentum and particle transport in materials. It is this aspect which
we will address in the present manuscript. We are going here to study
from a theoretical perspective the (hindered) Coulomb explosion of a
metal cluster embedded in a raregas matrix. That as such is a topic of
actual interest in cluster physics, see e.g
\cite{Teu01,Doe05b,Dit97a,Leb02,Der04a}. And it serves as a well
manageable test case for the more general issue of strong energy
impact on matter and subsequent transport processes which
are also of high current interest \cite{Che98,Tak01,Dou03}.

The theoretical description of embedded clusters is a great challenge
because of the inevitable complications by the cluster-matrix
interface and because of the huge number of atoms in the matrix. This
holds even more so for the simulation of their dynamics.  The quest
for robust and efficient schemes is omnipresent in all areas where one
deals with large compounds. The solutions rely usually on a
hierarchical modeling combining carefully quantum-mechanical and
classical descriptions, see e.g. \cite{Gre99a} for bio-molecules or
\cite{Mit93a,Roe01a} for chemical reactions on surfaces.  Along these
lines, we have developed a hierarchical model for the dynamics of
metal clusters in contact with rare gas materials
\cite{Ger04b,Feh05a}.  First successful tests of the
performance of this approach in the moderately non-linear regime were
presented in \cite{Feh05b,Dou06a}.

In this paper, we continue the study of the dynamics of embedded metal
clusters, considering now a violent laser excitation. The intensity is
chosen such that a free cluster would explode immediately while the
embedded configuration remains intact for a sufficiently long
observation time of several ps. As test case, we consider the dynamics
of a Na$_8$ cluster embedded in Ar$_{434}$, following irradiation by
an intense and short (33 fs) laser pulse. We will pay particular
attention to the impact of the environment on the dynamics of the
metal cluster and analyze the interlaced processes from several points
of view: detailed ionic and atomic trajectories, evolution of global
shape parameters (radius, quadrupole deformation) with their
associated vibrational time scales, and energy transfer providing
information on the various relaxation times.  Finally, we discuss a
possible identification by pump and probe analysis in relation to
recent experiments \cite{Sei00}.

\section{Formal framework}

In order to allow for a sufficiently large Ar matrix, we use a
hierarchical approach. The many different ingredients make a complete
representation rather bulky. In that section, we sketch only briefly
the basic constituents of the approach and refer to
\cite{Ger04b,Feh05c} for a detailed layout.

The Na cluster is treated in full microscopic
detail using quantum-mechanical single-particle wavefunctions
$\{\varphi_n({\bf r},t),n=1...N_{\rm el}\}$
for
the valence electrons. These are coupled non-adiabatically to classical
molecular dynamics (MD) for the Na ions
which are described by their positions
$\{{\bf R}_I,I=1...N_{\rm Na}\}$.
The electronic wavefunctions are propagated by time-dependent
local-density approximation (TDLDA).
The electron-ion interaction in the cluster is described by soft,
local pseudo-potentials. This TDLDA-MD has been validated for linear
and non-linear dynamics of free metal clusters \cite{Rei03a,Cal00}.

Two classical degrees-of-freedom are associated with each Ar atom:
center-of-mass
$\{{\bf{R}}_a,a=1...N_{\rm Ar}\}$, 
and electrical dipole moment which is parameterized as
$\{{\bf R}'_a={\bf R}_a+{\bf d}_a,a=1...N_{\rm Ar}\}$.
Note that the Ar dipole is practically handled by two constituents
with opposite charge, positive Ar core (at ${\bf R}_a$) and negative
Ar valence cloud (at ${\bf R}'_a$).
With the atomic dipoles, we explicitely treat the dynamical
polarizability of the atoms through polarization potentials
\cite{Dic58}. Smooth, Gaussian charge distributions are used for Ar
ionic cores and electron clouds in order to regularize the singular
dipole interaction.
We associate a Gaussian charge charge distribution to
both constituents having a width of the order of the 3p shell in Ar,
similar as was done in \cite{Dup96}. The Coulomb field of the
(softened) Ar dipoles provides the polarization potentials which
are the dominant agents at long ranges.
The Na$^+$ ions of the metal cluster have net
charge $q_{\rm Na}=+1$ and interact with the Ar dipoles predominantly
by the monopole moment. The small dipole polarizability of the Na$^+$
core is neglected.  The cluster electrons do also couple naturally to
the Coulomb field generated by the atoms.
%
%

The model is fully specified by giving the total energy of the system.
It is composed as
$$
  E_{\rm total}
  =
  E_{\rm Na cluster}
  +
  E_{\rm Ar}
  +
  E_{\rm coupl}
  +
  E_{\rm VdW}
  \quad.
$$
%
The energy of the Na cluster $E_{\rm Na cluster}$ consists out of
TDLDA for the electrons, MD for ions, and a coupling of
both by soft, local pseudo-potentials; that standard treatment is
well documented at many places, e.g. \cite{Rei03a,Cal00}.
The Ar system and its coupling to the clusters are described by
\begin{eqnarray}
\label{eq:ar_mat}
  E_{\rm Ar}
  =&&
  \sum_a \frac{{\bf P}_a^2}{2M_{\rm Ar}} 
  +
  \sum_a \frac{{{\bf P}'_{a}}^2}{2m_{\rm Ar}}
\nonumber\\
  &+&
  \frac{1}{2} k_{\rm Ar}\left({\bf R}'_{a}-{\bf R}_{a}\right)^2
\nonumber\\
  &+&  
  \sum_{a<a'}
  \Big[
    \int d{\bf r}\rho_{{\rm Ar},a}({\bf r})
    V^{\rm(pol)}_{{\rm Ar},a'}({\bf r})
\nonumber\\
   &&\qquad
    +
    V^{\rm(core)}_{\rm ArAr}({\bf R}_a - {\bf R}_{a'})
  \Big]
  \quad,
\\
\label{eq:ecoupl}
  E_{\rm coupl}
  =&&
  \sum_{I,a}\left[
    V^{\rm(pol)}_{{\rm Ar},a}({\bf R}_{I})
    +
    V'_{\rm NaAr}({\bf R}_I - {\bf R}_a)
  \right]
\nonumber\\
  &+&
  \int d{\bf r}\rho_{\rm el}({\bf r})\sum_a \Big[
    V^{\rm(pol)}_{{\rm Ar},a}({\bf r})
\nonumber\\
   &&\qquad\qquad
    +
    W_{\rm elAr}(|{\bf r}-{\bf R}_a|)
  \Big]
  \quad,
\\
\label{eq:Arpolpot}
  V^{\rm(pol)}_{{\rm Ar},a}({\bf r})
  &=&
  e^2{q_{\rm Ar}^{\mbox{}}}
  \Big[
   \frac{\mbox{erf}\left(|{\bf r}\!-\!{\bf R}^{\mbox{}}_a|
          /\sigma_{\rm Ar}^{\mbox{}}\right)}
        {|{\bf r}\!-\!{\bf R}^{\mbox{}}_a|}
\nonumber\\
   &&\qquad
   -
   \frac{\mbox{erf}\left(|{\bf r}\!-\!{\bf R}'_a|/\sigma_{\rm Ar}^{\mbox{}}\right)}
        {|{\bf r}\!-\!{\bf R}'_a|}
  \Big]
  \quad,
\\
\label{eq:VArel}
  W_{\rm elAr}(r)
  &=&
  e^2\frac{A_{\rm el}}{1+e^{\beta_{\rm el}(r - r_{\rm el})}}
  \quad,
\\
  V_{\rm ArAr}^{\rm (core)}(R)
  &=& 
  e^2 A_{\rm Ar}\Bigg[
  \left( \frac{R_{\rm Ar}}{R}\right)^{12}
 -\left( \frac{R_{\rm Ar}}{R}\right)^{6}
  \!\Bigg]
  \,,
\label{eq:VArAr}
\\
  V'_{\rm ArNa}(R)
  &=&
  e^2\Bigg[
  A_{\rm Na} \frac{e^{-\beta_{\rm Na} R}}{R}
\nonumber\\
   &-&
  \frac{2}{1\!+\!e^{\alpha_{\rm Na}/R}}
  \left(\frac{C_{\rm Na,6}}{R^6}\!+\!\frac{C_{\rm Na,8}}{R^8}\right)
  \Bigg]_,,
\label{eq:VpArNa}
\\
  &&
  4\pi \rho_{{\rm Ar},a}
  =
  \Delta V^{\rm(pol)}_{{\rm Ar},a}
  \quad,
\\
  E_{\rm VdW}
  &=&  
  e^2\frac{1}{2} \sum_a \alpha_a
  \Big[
    \frac{
       \left(\int{d{\bf r} \ {\bf f}_a({\bf r}) \ \rho_{\rm el}({\bf
  r})}\right)^2 }{N_{\rm el}}
\nonumber\\
   &&\qquad
      - \int{d{\bf r} \ {\bf f}_a({\bf r})^2 \rho_{\rm el}({\bf r})}
  \Big]
  \;,
\label{eq:EvdW}
\\
  &&
  {\bf f}_a({\bf r})
  =
  \nabla\frac{\mbox{erf}\left(|{\bf r}\!-\!{\bf R}^{\mbox{}}_a|
          /\sigma_{\rm Ar}^{\mbox{}}\right)}
        {|{\bf r}\!-\!{\bf R}^{\mbox{}}_a|}
  \quad,
\label{eq:effdip}
\\
  &&
  \mbox{erf}(r)
  = 
  \frac{2}{\sqrt{\pi}}\int_0^r dx\,e^{-x^2}
  \quad.
\end{eqnarray}
The calibration of the various contributions is taken from independent
sources, except perhaps for a final fine tuning to the NaAr dimer
modifying only the term $W_{\rm elAr}$ of Eq.(\ref{eq:VArel}). The
parameters are 
summarized in  table \ref{tab:params}. The third column of the table
indicates the source for the parameters. In the following, we report
briefly the motivations for the choices.

\begin{table*}
\begin{center}
\begin{tabular}{|l|l|l|}
\hline
 \rule[-8pt]{0pt}{22pt}
 $V^{\rm(pol)}_{{\rm Ar},a}$
&
 $q_{\rm Ar}
  =
  \frac{\alpha_{\rm Ar}m_{\rm el}\omega_0^2}{e^2}
 \;,\;
 k_{\rm Ar}
  =
  \frac{e^2q_{\rm Ar}^2}{\alpha_{\rm Ar}}
 \;,\;
 m_{\rm Ar}=q_{\rm Ar}m_{\rm el}$
&
 $\alpha_{\rm Ar}$=11.08$\,{\rm a}_0^3$
\\
 \rule[-12pt]{0pt}{22pt}
 &
 $\sigma_{\rm RG}
  =
  \left(\alpha_{\rm Ar}\frac{4\pi}{3(2\pi)^{3/2}}  \right)^{1/3}$
&
 \raisebox{12pt}{$\omega_0 = 1.755\,{\rm Ry}$}
\\
\hline
 \rule[-6pt]{0pt}{18pt}
 $W_{\rm elAr}$
&
 $A_{\rm el}=0.47  
 \;,\;
 \beta_{\rm el}$=1.6941\,/a$_0
 \;,\;
 r_{\rm el}$=2.2 a$_0$ 
&
 fit to NaAr
\\ 
\hline
 \rule[-8pt]{0pt}{22pt}
$V^{\rm(core)}_{\rm ArAr}$
&
 $A_{\rm Ar}$=$1.367*10^{-3}$ Ry
 $\;,\;
  R_{\rm Ar}$=6.501 a$_0$ 
&
fit to bulk Ar
\\
\hline
 \rule[-6pt]{0pt}{18pt}
 $V'_{\rm ArNa}$
&
 $\beta_{\rm Na}$= 1.7624 a$_0^{-1}
 \;,\;
 \alpha_{\rm Na}$= 1.815 a$_0
 \;,\;
 A_{\rm Na}$= 334.85 
&
\\
 \rule[-6pt]{0pt}{18pt}
&
 $C_{\rm{Na},6}$= 52.5 a$_0^6
 \;,\;
 C_{\rm Na,8}$= 1383 a$_0^8$
&
after \cite{Rez95}
\\
\hline
\end{tabular}
\end{center}
\caption{\label{tab:params}
Parameters for the various model potentials.
}
\end{table*}

Most important are the polarization potentials defined in
Eq.(\ref{eq:Arpolpot}). They are described by the
model of a valence electron cloud oscillating against the raregas core
ion. Its parameters are: 
$q_{\rm Ar}$ the effective charge of valence cloud, 
$m_{\rm Ar}=q_{\rm Ar}m_{\rm el}$ the effective mass of valence cloud, 
$k_{\rm Ar}$ the restoring force for dipoles, and 
$\sigma_{\rm Ar}$ the width of the core and valence clouds.
The $q_{\rm Ar}$ and $k_{\rm Ar}$ are adjusted to reproduce the dynamical
polarizability $\alpha_D(\omega)$ of the Ar atom at low frequencies,
i.e. we choose to reproduce the static limit 
$\alpha_D(\omega\!=\!0)$ and the second derivative
of $\alpha''_D(\omega\!''=\!0)$.
The width $\sigma_{\rm Ar}$ is determined consistently such that the
restoring force from the folded Coulomb force (for small
displacements) reproduces the spring constant $k_{\rm Ar}$.

The short range repulsion is provided by the various core potentials.
For the Ar-Ar core interaction, Eq.(\ref{eq:VArAr}), we employ a
Lennard-Jones type 
potential with parameters such that binding properties of bulk Ar are
reproduced.  The Na-Ar core potential, Eq.(\ref{eq:VpArNa}), is chosen
according to 
\cite{Rez95}. Note that the Na-Ar potential from \cite{Rez95} does
contain also a long range part $\propto\alpha_{\rm Ar}$ which accounts
for the dipole polarization-potential. We describe that long range part
explicitely and have to omit it here. We thus choose the form 
as given in $V'_{\rm ArNa}$.

The pseudo-potential $W_{\rm elAr}$,  Eq.(\ref{eq:VArel}), for the
electron-Ar core repulsion 
has been modeled according to the proposal of \cite{Dup96}. Its
parameters determine sensitively the binding properties of Na to the
Ar atoms.  We use them as a means for a final fine-tuning of the
model. The benchmark for adjustment is provided by the Na-Ar
dimer. The data (dimer binding energy, 
bond length, excitation energy) are taken from \cite{Gro98} and \cite{Rho02a}.
The adjustmenet shows some freedom in the choice of $A_{\rm el}$. We
exploit that to produce the softest reasonable core potential.

The Van-der-Waals energy $E_{\rm VdW}$,  Eq.(\ref{eq:EvdW}), is a
correlation from the 
dipole excitation in the Ar atom coupled with a dipole excitation in
the cluster. We exploit that $\omega_{\rm Mie}$ is far below the
excitations in the Ar atom.
This simplifies the term to the variance of the dipole operator in
the cluster, using again the regularized dipole operator ${\bf f}_a$
defined in Eq.~(\ref{eq:effdip}),
corresponding to the smoothened Ar charge distributions. The full
dipole variance is simplified in terms of the local variance.

The classical MD for Na ions, Ar atoms and Ar dipoles is
treated in full three dimensions. However,
the quantum mechanical treatment of the Na cluster involves two
approximations. 
The Kohn-Sham equations for the electron cloud involves axially
averaged potentials and the self-interaction correction (SIC) to TDLDA
is treated at the level of average-density SIC (ADSIC)
\cite{Leg02,And02b}. Both approximations enhance somewhat the barriers
for fragmentation of the cluster.
The axial approximation and ADSIC are needed to allow the large scale
calculations performed here. The SIC, on the other hand, is compulsory
for an appropriate description of ionization which is the leading
process in the initial stage of laser excitation.
These two approximations mean that we provide rather a qualitative
picture. The effects shown are certainly relevant. The thresholds
concerning charge state and laser field strength are probably not too
precise.
However we also dispose of a full three-dimensional treatment for the
valence electrons which, moreover, can handle various levels of SIC
and even exact exchange. We used that for occasional counterchecks and
we never found significant deviations from the approximate treatment.

Finally, a few words are in order about the possible range of validity for
that hierarchical model. The structure of the coupled systems is well
described by construction, see the predecessor in \cite{Dup96} and the
extensive testing in \cite{Ger04b}. The same holds for optical
response (see \cite{Feh05a,Ger04a}). However, one has to remain aware
that the model has limitations with respect to allowed frequencies and
amplitudes. The dynamical response of the Ar atoms is tuned to
frequencies safely below the Ar resonance peak, i.e. safely below 1
Ry. This is well fulfilled in all our calculations because the leading
frequency is the cluster's Mie plasmon at around 0.2 Ry.  The
amplitudes of the dipole oscillations in Ar should stay below the
threshold where free electrons are released from the Ar atom into the
matrix. We have checked that process by fully quantum mechanical
calculations of laser excitation on Ar atoms and we find a critical
field strength of order of 0.1 Ry/a$_0$. We protocol during our
calculation the actual field strength at each Ar site and thus check
that we do not exceed that limit in our simulations.

\section{The test case}

We focus here on a Na$_8$ cluster embedded in a large Ar cluster with
434 atoms. This should serve as a finite model of an Ar matrix,
although aspects of finite mixed clusters will also be addressed
briefly. From a more general perspective, we encounter here a typical
example of a chromophore inside an inert environment. The case is
particularly interesting  because the mix of simple materials allows
to concentrate on the basic effects of a chromophore without the need
to bother about the elaborate technical details of more complex
materials, as e.g. large organic molecules. 
We consider the case of low initial temperatures, safely
below the Ar melting point of 84 K, where spatial fluctuations of the
atoms and ions can be safely neglected.

The small Na cluster at low temperature has a very clean excitation
spectrum with rather narrow excitation peaks in the
plasmon range around 2.5 eV \cite{Rei03a}.
Correlation effects beyond TDLDA remain negligible 
for the global observables considered here.
Spectral fragmentation from coupling to energetically close
electron-hole states (Landau fragmentation) is negligible for
that system size because the Mie surface plasmon resides
in a gap of such states \cite{Rei96c}.
Broadening from higher correlations is also small
\cite{Bon96a}, again due to the spectral gap for that magic
electron number.
Thermal broadening in the initial state is negligible
at the presently studied low temperatures  \cite{Ell95}.
Larger electronic temperatures develop in the course of the dynamics
and they call, in principle, for a description beyond pure mean
field. Electron-electron collisions could be included by switching to
a semi-classical Vlasov-Uehling-Uhlenbeck description of the electron
cloud \cite{Feh05c,Gig01a}. But long-time semi-classical propagation
is still extremely hard to stabilize and, as we will see in
section~\ref{sec:energ}, electrons are thermally almost decoupled from
ions. Thus we can neglect that effect for time scales explored in the
present study.  
The surrounding Ar cluster has to be sufficiently large to absorb the
energy from the highly excited chromophore without breakup and to
avoid specific size or surface effects.  We have checked matrix sizes
from 30 to 1048 Ar atoms for the dynamical regime intended here and
found qualitatively similar behaviors from 164 Ar atoms on. To be
on the safe side, the present results have been obtained for a
model matrix of 434 Ar atoms. 
The setup was generated by cutting a closed pack of 447 Ar atoms out
of a crystal, removing further 13 atoms from the center, placing the
Na$_8$ in the cavity, and finally re-optimizing the whole
configuration by Monte-Carlo simulated annealing techniques. The
Na$_8$ cluster with a 
radius of 7 a$_0$ (r.m.s.-radius of 5.5 a$_0$) then resides in a cavity
of about 10 a$_0$ radius.  The outermost Ar shell has a radius of 31
a$_0$. This provides a substantial coverage of the embedded Na$_8$.
The spatial structure is still very close to the configuration before
re-optimization \cite{Feh05a}. The Na$_8$ consists out of two rings of
four ions each and the rings have a relative rotation of 45$^\circ$ to
minimize the Coulomb repulsion between the ions. The global shape is
axially symmetric, only slightly oblate deformed.  The surrounding Ar
maintains basically its original crystal structure with some small
rearrangement for the innermost shells.

The dynamics in the linear regime was studied in \cite{Feh05a}.  A
first case of moderate laser excitations was discussed in
\cite{Feh05b}.  Here we aim at a more violent, still non-destructive,
process. To that end, we irradiate the Na$_8$@Ar$_{434}$ system with a
laser of frequency 1.9 eV, intensity $2.4*10^{12}$ W/cm$^2$, and a
cos$^2$ pulse envelope with total pulse length of 100 fs (FWHM = 33
fs).  The laser is polarized along the symmetry axis of the system,
henceforth called $z$ axis. The rather short and intense laser
excitation produces very quickly a net ionization stage $3^+$ and
deposits a large amount of energy in the system.  We then continue to
propagate the system up to about 10 ps in order to cover the relevant
time scales. 

\section{Results and discussion}

\subsection{Detailed view at the propagation of ions and atoms}

\begin{figure*}
\centerline{\includegraphics[width=13.2cm,angle=0]{na8_ar434_o14_e0166_coord.eps}
}
\caption{\label{fig:pos} 
Time evolution of the the atomic (full lines) and ionic (dotted lines)
$z$-coordinates (lower panel) and radial distances
$r=\sqrt{x^2+y^2+z^2}$ (upper panel) for Na$_8$\@Ar$_{434}$ excited
with a laser of frequency $\omega=1.9$ eV, intensity $2.4*10^{12}$
W/cm$^2$, and a cos$^2$ pulse profile with FWHM = 33 fs. 
The laser was polarized along the $z$ axis which is also
the symmetry axis of the system.
}
\end{figure*}

The initial reaction of the system is dominated by the large dipole
response of the electrons. For the short pulses used here, the strong
electrical field with subsequently large dipole amplitude
leads to a direct emission of $~$3 electrons which
escape instantaneously (i.e. within 3-10 fs).
%
%
The thus produced large Coulomb pressure leads to a
rearrangement of the whole system (Na ions and Ar atoms) at
ionic/atomic time scale as is demonstrated in figure \ref{fig:pos}
showing the time evolution of Na ions and Ar atoms in detail.
It exhibits different interesting 
time scales. 
%
The first time scale (after the electronic one, of order of a few fs) is
observed in the reaction of the Na ions.  They perform a Coulomb
explosion whose first signs can be seen at about 50 fs. That proceeds
almost identical to the similar case of a free Na$_8$ cluster lifted
to charge 3$^+$. But the ``explosion'' is abruptly stopped at about
200 fs
when the ions hit
the repulsive core of the first shell of Ar. The ionic motion
turns to damped oscillations around a (r.m.s.) radius of about 7 a$_0$. 
In atoms, one also observes two stages, although on longer
time scales. The first phase is a spreading of the
perturbation (caused by the stopping of Na ions) into the various
Ar shells. This is especially visible along the $z$ axis which allows
to read off the propagation speed of this perturbation as 20-30
a$_0$/ps.
We have estimated the sound velocity in the corresponding large pure
Ar cluster (Ar$_{447}$) by computing its vibrational spectrum. The
radial compression mode corresponds to the sound mode in bulk
material.  Its frequency is $\omega_{\rm vib}=1.8$ meV.  The momentum
of the radial wave is $q=\pi/R$ where $R=30$ a$_0$ is the cluster
radius. The sound velocity is then estimated as $v_{\rm
sound}=\omega_{\rm vib}/q\approx30$ a$_0$/ps which is very close to
the propagation speed as observed in the figure.
This suggests an interpretation as a sound wave sent by the
initial bounce of the Na ions.

Once  transferred to a given Ar shell the perturbation
generates oscillations combined with some diffusion which
after about 1.5 ps has spread over all shells.
The perturbations affect even
the outermost shell which thus acquires some oscillations.
That is an effect of finite size of the ``matrix''.  The oscillation
amplitude shrinks with increasing system size and grows towards
smaller matrices.  For example, it was found to be smaller for the
larger test case with 1048 Ar atoms and the last Ar atoms are even
kicked away for matrices at and below 164 Ar atoms. The reason is
twofold: smaller system provide, first, less Ar shells to absorb
energy, and second, less dipole binding as whole.  This question will
be discussed more systematically in a larger forthcoming paper.

The relaxation of matrix oscillations is much slower than that of
Na ions and beyond the time scale computed here.  These long time
scales for full relaxation and evaporation of Ar atoms are well known
from experiments of dimer molecules embedded in Ar clusters, see
e.g. \cite{Vor96a}.

We finally comment briefly on the finite net charge of our test
system. It emerges because the electrons propagate almost unhindered
through the surrounding Ar cluster and eventually escape to
infinity. In a macroscopically large matrix, the electrons would be
stuck somewhere within the range of their mean free path, drift very
slowly back towards the now attractive cluster well, and eventually
recombine there. We have simulated that by using reflecting boundary
conditions rather than absorbing ones and we find recombination times
of the order of several 10 ps. Furthermore the ionic/atomic
rearrangements following the irradiation turn out to be qualitatively
very similar whatever the boundary conditions. Thus the present
scenario should provide a pertinent picture for several ps, the time
window studied here.

\subsection{Time evolution of global shapes}

\begin{figure*}
\centerline{\includegraphics[width=13.2cm]{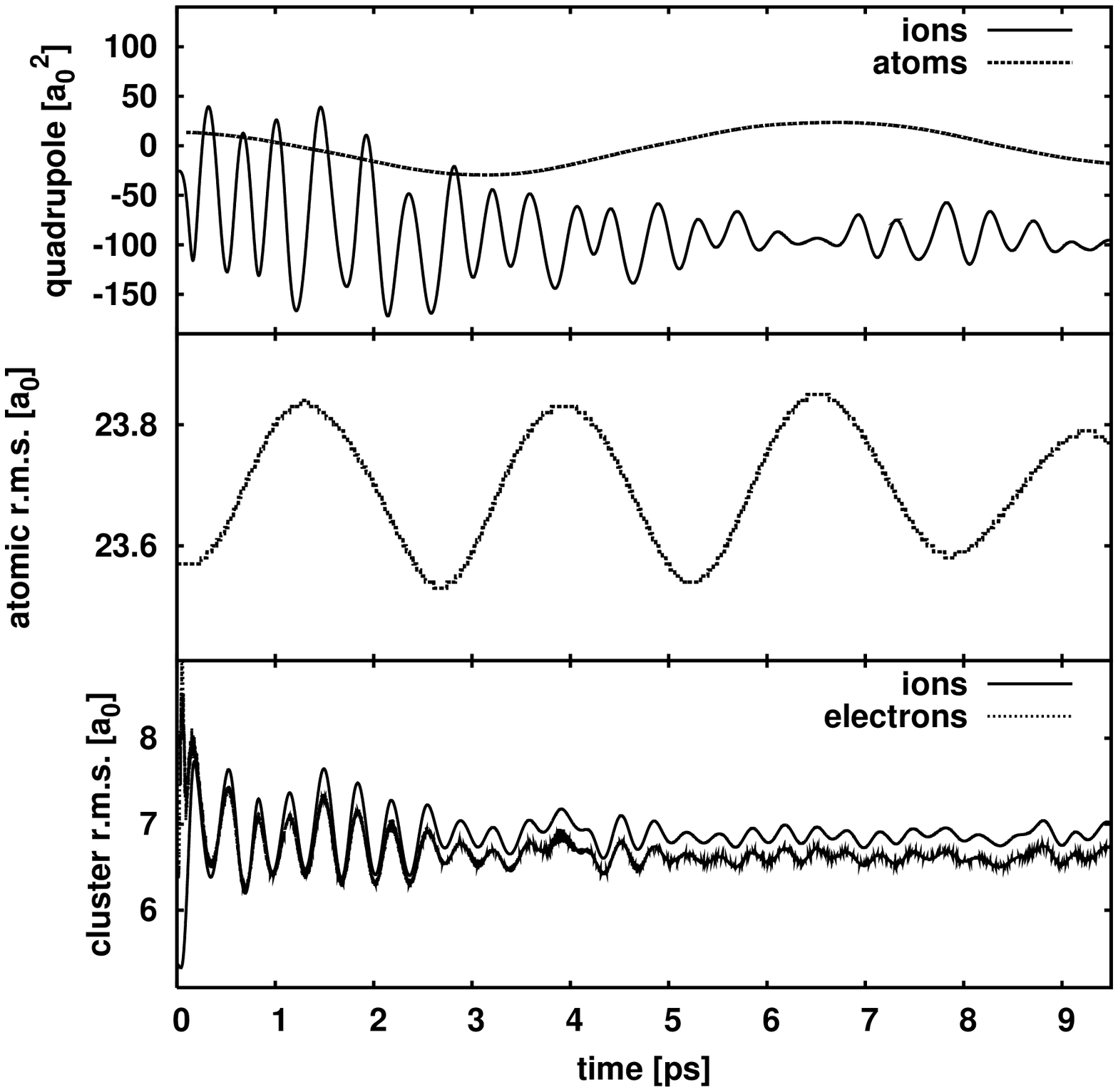}}
\caption{\label{fig:shapes}
Time evolution of global shape observables of atoms (solid lines),
ions (dotted lines), and electrons (fine dashed lines).
Lowest panel:  r.m.s. radii of cluster constituents, ions and
electrons.
Middle panel: r.m.s. radius of Ar atoms.
Upper panel: Quadrupole moments of ions and atoms.
}
\end{figure*}
Figure \ref{fig:shapes} visualizes the ionic/atomic dynamics in terms of
simple global observables. The middle panel shows the ionic
r.m.s. radius which provides an even clearer picture of the
ionic dynamics described above. 
Let us first concentrate on the 
the lower two panels showing the global
radial oscillations (breathing). 
The electronic radius expands during the short excitation
phase. However, as soon as the laser is switched off after 100 fs,
the electronic radius quickly relaxes towards the ionic one, as can be
seen in the lower panel of figure~\ref{fig:shapes}. Indeed
the strong monopole
Coulomb forces ties electronic and ionic radius always closely
together. The radial oscillations of the cluster have the typical
cycle of about 250 fs, well known from free Na clusters \cite{Rei02d}
and re-established for embedded ones \cite{Feh05b}. The most
interesting aspect is here the clearly visible relaxation of the
cluster oscillations at a time scale of about 4 ps. It seems that a
slow energy transfer from the ionic cluster vibrations into the
raregas environment takes place. This will be inspected further in
section \ref{sec:energ}. 
The breathing oscillation of the Ar environment (i.e. of the atomic
radii) proceeds at a cycle of about 3 ps (see middle panel of
figure~\ref{fig:shapes}),  corresponding to a volume
frequency of about 1.4 meV.
This is within factor two the surface vibration frequency observed in
neutral Ar clusters of that size \cite{Buc94a}, which is quite
gratifying
in view of the structural differences (Na cluster embedded,
high net charge) between these two systems. 
A detailed comparison will have to be worked out in future studies.

The upper panel complements the analysis by the quadrupole moments of
the Na cluster (ions) and of the Ar atoms. The quadrupole oscillations
of both subsystems are again at their very own time scale which a very
long cycle time, about 7 ps, for quadrupole motion of the Ar atoms.
The Na cluster shows, moreover, a global trend to finally oblate
shapes.
Note that a similar effect, the build-up of persistent quadrupole
deformations following laser irradiation, has been observed for Ag
clusters embedded in glass \cite{Sei00}.  Glass is a much harder
material than Ar but we may speculate that the strong laser pulse
provides also an intense heating, and thus melting, of the immediate
neighborhood of the embedded cluster. For then the mechanisms of
deformation may be similar.

\subsection{Energetic aspects}
\label{sec:energ}

\begin{figure*}
\centerline{\includegraphics[width=13.0cm]{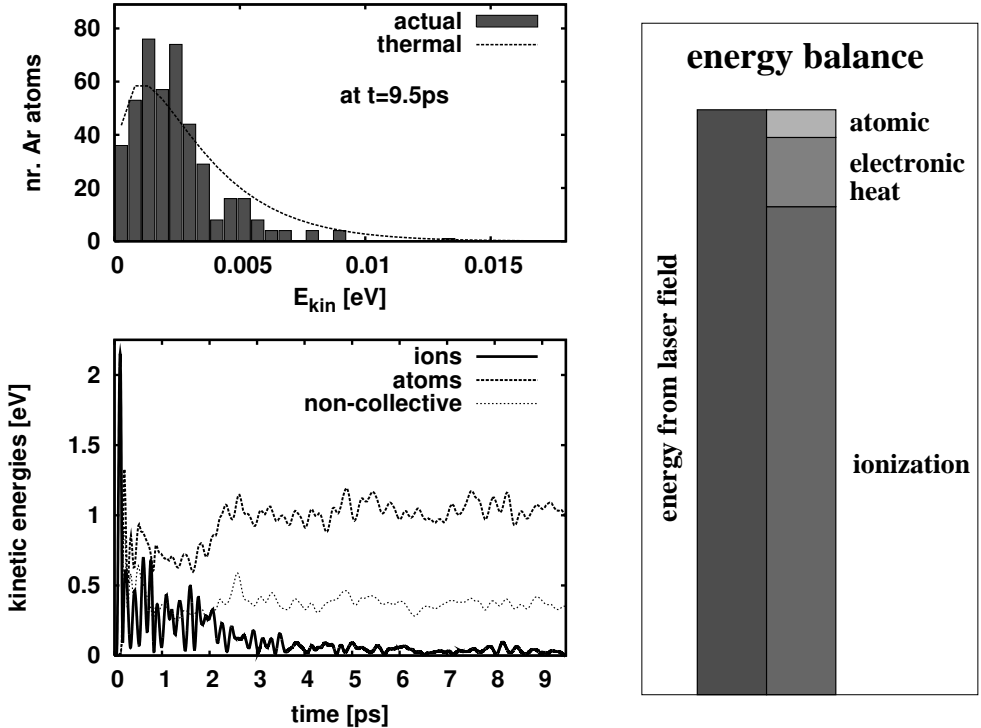}}
\caption{\label{fig:therm}
Lower left: time evolution of kinetic energies for atoms and ions;
the line denoted ``non-collective'' shows the internal kinetic
energy of the atoms after subtracting the contribution from 
global radial breathing. 
Upper left:
Distribution of atomic kinetic energies at time $t=9.5$ ps
where the boxes show the actual atomic distribution sampled in energy
bins and the 
dotted line stands for the equivalent thermal
(Maxwellian) distribution with the temperature as given by
the instantaneous kinetic energy.
Right:
energy balance, i.e. input through the laser field (left)
and its distribution over various subsystems (right).
}
\end{figure*}
The lower left panel of figure \ref{fig:therm} shows the kinetic
energies of ions and atoms. The Na ions gather energy very quickly
from the starting Coulomb explosion. At around 200 fs, the quickly
acquired energy is transferred abruptly to the atoms, leaving only one
fifth of kinetic energy for the first phase of ionic
oscillations. From then on, both kinetic energies proceed for a while
at nearly constant values with some residual oscillations.  
At around 2 ps, a slow transition seems to take place where the ionic
energy flows towards the atomic subsystem. The situation stabilizes after
3 ps to a thermal share of energies where practically all kinetic
energy resides in the atoms and only a tiny fraction remains for
the ions. 
However, the notion of ``thermal share'' does not imply a full 
equilibrium  even at final simulation time of 9.5 ps.
We know from figure \ref{fig:shapes} that there are very
slow and long lasting atomic vibrations. We should call the present
situation more correctly a local (or instantaneous) equilibrium.

The global atomic breathing oscillations carry still a large amount of
energy. To quantify that aspect we
 have computed the collective energy. The fine dashed line
in the lower left panel shows the remaining ``internal'' kinetic energy
of the atoms. And indeed, almost two thirds of the energy reside still
in the collective breathing yet awaiting final thermalization at a time
scale far beyond 10 ps.
Nonetheless, it is interesting to have a look at the distribution of
kinetic energies. This is shown in the upper left panel
of figure \ref{fig:therm}. It is the
distribution of the total kinetic energies and it is compared with 
the thermal distribution at a temperature of 20 K, equivalent
to the full kinetic energy at the late times in the plot.
Both distributions compare fairly well in view of the fact
that full equilibrium has not yet been reached (thus allowing for
somewhat larger fluctuations).

The right panel sketches the main energy share in the reaction.  The
system acquires as much as 23 eV energy from the laser pulse.  The by
far largest portion, namely 83\%, is carried away very quickly by the
emitted electrons. Two third from what remains flow into ionic and
atomic motion. We have learned from the lower left panel that this
energy arises from the stopped Coulomb explosion, and is shared in two
steps between ions  and atoms: initially, the ions acquire a large
potential energy from sudden charging, this is converted through
``Coulomb explosion'' instantaneously to a large kinetic
energy, which then is transferred quickly to the atoms
when the ``explosion'' is abruptly stopped. A
somewhat slower second phase of energy transfer follows after 2 ps
which is related to rearrangement and some exchange with still
available potential energy.
One third of the remaining energy, i.e. about 1 eV, remains with the
electrons as internal kinetic energy (this is the excess kinetic
energy beyond the unavoidable quantum mechanical minimum, see
\cite{Cal00}). That is very little in view of the fact that it has
been the cluster's electron cloud which served as doorway to absorb all
energy from the laser field. It is, however, a great deal in terms of
temperature. The electronic temperature corresponds to 1500 K whereas
the ionic/atomic temperature has yet come up to only 18 K. That makes
it clear that the process resides still in a transient stage where the
electronic and the ionic/atomic subsystems have reached a local
equilibrium amongst themselves. Full equilibration and subsequent
evaporative cooling will follow at a much longer time scale, far
beyond the capabilities of the very detailed description through a
TDLDA-MD approach.

There remains the question which energetic aspects are generic for
metal clusters in a rare gas environment and which are specific to the
particular mixed system studied here. The generic aspects are related
to what happens with the cluster and its immediate rare gas vicinity.
This concerns the energy absorption from the laser which proceeds
predominantly through the metal cluster's electrons, the stopping of
the Coulomb explosion and associated fast energy transfer to the
atoms, the relaxation time scales in these early phases of a few ps,
and the energy balance in absolute numbers. The particular property
which depends on the matrix size is most of all the atomic
temperature. We find that the same absolute amount of energy is
transferred from the laser to the Na$_8$ cluster, independent of
matrix size. The energy loss through ionization is also the same.
Thus the final atomic temperature will then be inversely proportional
to the matrix size.

\subsection{Possible pathways to experimental analysis}


The question remains how one could analyze the key pattern of the
embedded cluster dynamics experimentally.
The most interesting effect is the hindered explosion of the
imprisoned Na cluster and its subsequent shape oscillations in the Ar
cavity.  Such shape oscillations and eventual relaxation to deformed
shape have been produced and observed experimentally for Ag clusters
embedded in glass \cite{Sei00} or deposited on a substrate
\cite{Wen99}. The (time-dependent) cluster deformation is verified by probe
pulses measuring the actual optical response of the metal cluster,
exploiting the pronounced features of the Mie plasmon resonance.
Indeed, the plasmon couples strongly to light at very specific
frequencies and these plasmon frequencies are uniquely related to the
cluster shape. 
For free clusters, we have discussed a rather simple
pump-and-probe setup which allows to map in a
unique fashion the evolution of radial shape \cite{And02}, of
quadrupole deformations \cite{And04} and of elongation in a fission
process~\cite{Din05,And06}.
We have computed that for the case
considered here. To this end, we take the instantaneous configuration at a given
time and compute the optical response in all three spatial directions.  
These expensive calculations have been performed in full 3D TDLDA
(which allows to properly access the directions perpendicular to the
laser polarization axis) on a representative time of extremal
deformations.  The other points are interpolated by well benchmarked
plasmon calculations for a smaller subsystem.
%
\begin{figure}
\centerline{\includegraphics[width=7.8cm]{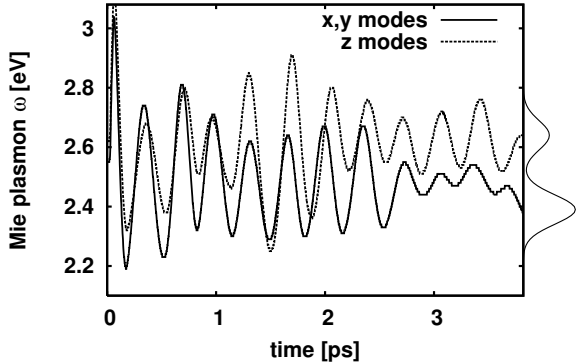}}
\caption{\label{fig:opt} 
Time evolution of the estimated Mie plasmon
frequencies in $x$, $y$ and $z$ directions for the embedded Na$_8$
cluster. The cluster remains always almost axially symmetric such
that the $x$ and $y$ modes are nearly degenerate. The spectral
distribution at the last time point is indicated to the right
of the figure.}
\end{figure}
Results are shown in figure \ref{fig:opt}. The radial oscillations
(middle panel of figure \ref{fig:shapes}) relate to the oscillations
of the average peak position which are particularly well visible here
in the early phase up to about 1.5 ps. The then evolving oblate
deformation leads to a splitting of the resonance peak where the
shorter extension along $z$ is associated to a blue shift of the mode
and the larger extension in orthogonal direction to a red shift.  The
$x$ and $y$ modes are nearly degenerate due to near axial symmetry of
the total system. Pump and probe analysis of the time-dependent
plasmon peaks thus provides worthwhile information about the shapes of
the embedded cluster, which can be measured by probe pulse as
explained in \cite{And02,And04,Din05,And06}.  Note here that the experimental
results obtained for Ag clusters in glass \cite{Sei00} are in qualitative
agreement with our findings. Our model thus provides a microscopic
interpretation, the first one, to the best of our knowledge, of these
experimental investigations.

\section{Conclusion}

In this paper, we have analyzed the dynamics of a Na cluster embedded
in an Ar matrix following irradiation by an intense laser pulse.  We
have used a hierarchical model with a detailed quantum-mechanical
treatment of the electron cloud in the Na cluster and a classical
description of the Na ions and surrounding Ar atoms whereby we include
the Ar dipole moments to account for the dynamical
polarizabilities. As test case, we have considered Na$_8@$Ar$_{434}$
which provides several interesting aspects~: It is a prototype of a
chromophore in an else-wise inert material, it serves as model for a
cluster embedded in a raregas matrix, and we do also find interesting
features which are specific for a finite mixed cluster.  For the
excitation mechanism, we consider irradiation by a short 33 fs laser
pulse with intensity of about $10^{12}$W/cm$^2$.  The laser couples
almost exclusively to the clusters electron cloud and deposits about 23
eV energy in the system. This leads to immediate ionization of Na$_8$
to a 3$^+$ charge state. The subsequent Coulomb explosion of the
Na$_8$ is abruptly stopped by the Ar atoms, turning the motion into
steady shape oscillations of the cluster while the absorbed momentum
spreads in a radial sound wave propagating through the atoms and
triggers global radial oscillations thereof. A second phase
around 2-3 ps transfers most of the remaining ionic energy to
the atoms and cools the ions to almost atomic temperature. The
relaxation process can be nicely seen in oscillations of cluster
radius and deformation. Both these observables are accessible to
experimental observation by an especially designed pump-and-probe
setup.

A further interesting feature is that the Ar surroundings do not only
stop the Coulomb explosion. They even stabilize the high charge state
at a time scale which is sufficiently long for an observation of ionic
and atomic motion. This is probably a particular feature of finite
mixed clusters. The surrounding Ar system should be large enough to
deliver stabilization through induced dipole attraction, but small
enough such that the cluster electrons which are directly emitted due
to the laser excitation can travel nearly unhindered through the Ar
shells and finally escape to whole system. The degree of charge
stabilization in dependence of system size and the mean free path of
electrons through ordered raregas shells is a task for future studies.

\bigskip

Acknowledgments: 
This work was supported by the DFG (RE 322/10-1),
the french-german exchange program PROCOPE,  
 the CNRS Programme "Mat\'eriaux" 
  (CPR-ISMIR), Institut Universitaire de France, the
  Humboldt foundation and a Gay-Lussac prize.

%


%
%


\bigskip


\begin{thebibliography}{10}

\bibitem{Fey60}
R. Feynman, Engineering and Science {\bf 23}, 22 (1960). 

\bibitem{Ebe02}
W. Eberhardt, Surf. Sci. {\bf 500}, 242 (2002).

\bibitem{Kre93}
U.~Kreibig and M.~Vollmer,
 {\em Optical Properties of Metal Clusters}, vol.~25.
 Springer Series in Materials Science, 1993.

\bibitem{Hee93}
Walt.~A. de~Heer,
 { Rev. Mod. Phys.} {\bf 65}, 611 (1993).

\bibitem{Bra93}
M.~Brack,
 { Rev. Mod. Phys.} {\bf 65}, 677 (1993).

\bibitem{Hab94ab}
H.~Haberland, editor,
 {\em { Clusters of Atoms and Molecules 1 and 2}}, vol.~52 and 56.
 Springer Series in Chemical Physics, Berlin, 1994.


\bibitem{Eka99}
W.~Ekardt, edt.,
 {\em {Metal Clusters}}.
 Wiley, New York, 1999.

\bibitem{Rei03a}
P.-G. Reinhard and E.~Suraud,
 {\em {Introduction to Cluster Dynamics}}.
 Wiley, New York, 2003.

\bibitem{Cal00}
F.~Calvayrac, P.-G.~Reinhard, E.~Suraud, and C.~A. Ullrich,
 { Phys. Rep.} {\bf 337}, 93 (2000).

\bibitem{Teu01}
S.~Teuber, T.~D\"oppner, T.~Fennel, J.~Tiggesb\"aumker, and K.~H. Meiwes-Broer,
 { Eur. Phys. J. D} {\bf  16}, 59 (2001).

\bibitem{Bes99}
B.~Bescos, B.~Lang, J.~Weiner, V.~Weiss, E.~Wiedemann, and G.~Gerber,
 { Euro. Phys. J. D} {\bf  9}, 399 (1999).

\bibitem{And02}
K.~Andrae, P.-G.~Reinhard, and E.~Suraud,
 { J. Phys. B} {\bf  35}, 1 (2002).

\bibitem{Doe05b}
T.~D\"oppner, Th.~Fennel, Th.~Diederich, and J.~Tiggesb\"aumker 
and K.H Meiwes-Broer,
 { Phys. Rev. Lett.} {\bf  94}, 013401 (2005).

\bibitem{Din05}
P.M.~Dinh, P.-G.~Reinhard, and E.~Suraud,
J. Phys. B {\bf 38}, 1637 (2005).

\bibitem{And06}
K.~Andrae, P.M.~Dinh, P.-G.~Reinhard, and E.~ Suraud
Comp. Mat. Sci.	 {\bf 35}, 169 (2006).


\bibitem{Dit96}
T.~Ditmire, T.~Donnelly, A.~M. Rubenchik, R.~W Falcone, and M.~D. Perry,
 { Phys. Rev. A} {\bf  53}, 3379 (1996).

\bibitem{Buz96}
S.~A. Buzza, E.~M. Snyder, D.~A. Card, D.~E. Folmer, and A.~W.~Castleman Jr,
 { J. Chem. Phys.} {\bf  105}, 7425 (1996).
\bibitem{Rei04r}
P.--G. Reinhard, E. Suraud,
Encycl. Nanosc. Nanotechn. {\bf 2}, 717 (2004)

\bibitem{Bel04r}
M. Belkacem, M. A. Bouchenne, P.--G. Reinhard, and E. Suraud,
Encycl. Nanosc. Nanotechn. {\bf 8}, 575 (2004)


\bibitem{Goe95}
T.~G\"otz, F.~Tr\"ager, M.~Buck, C.~Dressler, and F.~Eisert,
 { Appl. Phys. A} {\bf  60}, 607 (1995).

\bibitem{Bal00}
F.~Balzer, S.~D. Jett, and H.-G. Rubahn,
 { Solid Films} {\bf  372}, 78 (2000).

\bibitem{Koh00}
C.~Kohl, P.-G. Reinhard, and E.~Suraud,
 { Euro. Phys. J. D} {\bf  11}, 115 (2000).

\bibitem{Sei00}
G.~Seifert, M.~Kaempfe, K.-J. Berg, and H.~Graener,
 { Appl. Phys. B} {\bf  71}, 795 (2000).

\bibitem{Oua05a}
H.~Ouacha, C.~Hendrich, F.~Hubenthal, and F.~Tr\"ager,
 { Appl. Phys. B} {\bf  81}, 663 (2005).



\bibitem{Bin01}
C. Binns,
{Surf. Sci. Rep.} {\bf 44}, 1 (2001).


\bibitem{Mil99}
P. Milani, S. Iannotta,
{\em Cluster Beam Synthesis of Nanostructured Materials},
Springer, Berlin 1999.

\bibitem{Nil00}
N.~Nilius, N.~Ernst, and H.-J. Freund,
 { Phys. Rev. Lett.} {\bf  84}, 3994 (2000).

\bibitem{Leh00}
J.~Lehmann, M.~Merschdorf, W.~Pfeiffer, A.~Thon, S.~Voll, and G.~Gerber,
 { Phys. Rev. Lett.} {\bf  85}, 2921 (2000).

\bibitem{Gau01}
M.~Gaudry, J.~Lerm\'e, E.~Cottancin, M.~Pellarin, J.-L. Vialle, 
M.~Broyer, B.~Pr\'evel,
 { Phys. Rev. B} {\bf  64}, 085407 (2001).

\bibitem{Die02}
T.~Diederich, J.~Tiggesb\"aumker, and K.~H. Meiwes-Broer,
 { J. Chem. Phys.} {\bf  116}, 3263 (2002).

\bibitem{May01}
C.~Mayer, R.~Palkovits, G.~Bauer, and T.~Schalkhammer,
 { J. Nanoparticle Res.} {\bf  3}, 361 (2001).

\bibitem{Dub02}
B.~Dubertret, P.~Skourides, D.~J Norris, V.~Noireaux, A.~H. Brivanlou, and
A.~Libchaber,
 { Science} {\bf  298}, 1759 (2002).

\bibitem{Bar02b}
M.~Bargheer, M.~Guhr, and N.~Schwentner,
 { J. Chem. Phys} {\bf  117}, 5 (2002).

\bibitem{Niv00a}
M.~Y. Niv, M.~Bargheer, and R.~B. Gerber,
 { J. Chem. Phys} {\bf  113}, 6660 (2000).


\bibitem{Dit97a}
T. Ditmire,  J. W. G. Tisch,  E. Springate,  M. B.  Mason, 
  N. Hay,  R. A. Smith, and J. Marangos and M. H. R. Hutchinson,
{Nature} {\bf 386}, 54 (1997).


\bibitem{Leb02}
{M. A. Lebeault,  J. Viallon,  J. Chevaleyre,  
                  C. Ellert,  D. Normand ,  M. Schmidt,  
                  O. Sublemontier,  C. Guet, and B. Huber},
{Eur. Phys. J. D} {\bf 20}, 233 (2002).

\bibitem{Der04a}
{T. E. Dermota, Q. Zhong, and A. W. Castleman},
{Chem. Rev.} {\bf 104}, 1861 (2004).

\bibitem{Che98}
H.-P. Cheng and J. D. Gillaspy,
{Comp. Mat. Sci.} {\bf 9}, 285 (1998).

\bibitem{Tak01}
{S. Takami, K. Suzuki, M. Kubo, and A, Miyamoto},
{J. Nanoparticle Res.} {\bf 3}, 213 (2001).

\bibitem{Dou03}
Y. Dou, N. Winograd, B. J. Garrison, and L. V. Zhigilei,
{J. Phys. Chem. B} {\bf 107}, 2362 (2003).



\bibitem{Gre99a}
N. Gresh, O. Parisel, and C. Giessner-Prettre,
THEOCHEM {\bf 458}, 27 (1999).



\bibitem{Mit93a}
{P. J. Mitchell and D. Fincham},
J. Phys. CM {\bf 5}, 1031 (1993).


\bibitem{Roe01a}
{V. A. Nasluzov, K. Neyman,  U. Birkenheuer, and N. Rösch},
J. Chem. Phys. {\bf 115}, 17 (2001).

\bibitem{Ger04b}
B.~Gervais, E.~Giglio, E..~Jaquet, A..~Ipatov, P.-G. Reinhard, and E.~Suraud,
 { J. Chem. Phys.} {\bf  121}, 8466 (2004).

\bibitem{Feh05a}
F.~Fehrer, P.-G.~Reinhard, E.~Suraud, E.~Giglio, B.~Gervais, and A.~Ipatov,
 { Appl. Phys. A} {\bf  82}, 152 (2005).

\bibitem{Feh05b}
F.~Fehrer, P.-G.~Reinhard, and E.~Suraud,
 { Appl. Phys. A} {\bf  82}, 145 (2005).

\bibitem{Dou06a}
J.~Douady, B.~Gervais, E.~Giglio, A.~Ipatov, and E.~Jacquet,
 { J. Mol. Struct.} {\bf 786}, 118 2006.

\bibitem{Feh05c}
F. Fehrer, M. Mundt, P.--G. Reinhard, and E. Suraud,
Ann. Phys. (Leipzig) {\bf 14}, 411 (2005).


\bibitem{Dic58}
B.~G. Dick and A.~W. Overhauser,
 { Phys. Rev.} {\bf  112}, 90 (1958).

\bibitem{Rez95}
G.~Reza Ahmadi, J.~Alml\"of, and J.~R\/oegen,
{Chem. Phys.} {\bf 199}, 33 (1995).

\bibitem{Dup96}
F.~Dupl\`ae and F.~Spiegelmann,
 { J. Chem. Phys.} {\bf  105}, 1492 (1996).


\bibitem{Ell95}
C.~Ellert, M.~Schmidt, C.~Schmitt, T.~Reiners, and H.~Haberland,
 { Phys. Rev. Lett.} {\bf  75}, 1731 (1995).

\bibitem{Rei96c}
P.-G. Reinhard, O. Genzken, and M. Brack,
Ann. Phys. (Leipzig) {\bf 5}, 576 (1996).


\bibitem{Bon96a}
V. Bonacic-Koutecky, J. Pittner, C. Fuchs, P. Fantucci,
M. F. Guest, and J. Koutecky,
J. Chem. Phys. {\bf 104}, 1427 (1996).



\bibitem{Vor96a}
V.~Vorsa, P.~J.~Campagnola, S.~Nandi, M.~Larsson, and  W.~C.~Lineberger,
{J. Chem. Phys.} {\bf 105}, 2298 (1996).



\bibitem{Rei02d}
P.-G. Reinhard, E. Suraud,
Eur. Phys. J. D {\bf 21}, 315 (2002)


\bibitem{Buc94a}
U.~Buck and R.~Krohne, 
{Phys. Rev. Lett.} {\bf 73}, 947 (1994).

 
\bibitem{Wen99}
T. Wenzel, J. Bosbach, A. Goldmann, and F. Tr\"ager,
Appl. Phys. B {\bf 69}, 513 (1999)

\bibitem{And04}
K.~Andrae, P.-G.~Reinhard, and E.~Suraud,
 { Phys. Rev. Lett.} {\bf  92}, 173402 (2004).
\bibitem{Gro98}
M.~Gross and F.~Spiegelmann.
\newblock {J. Chem. Phys.} {\bf 108}, 4148 (1998).

\bibitem{Rho02a}
M.~B. {El Hadj Rhouma}, H.~Berriche, Z.~B. Lakhdar, and F.~Spiegelman.
\newblock {J. Chem. Phys.}, {\bf 116}, 1839 (2002).


\bibitem{Leg02},
C.~Legrand, E.~Suraud, and P.-G.~Reinhard.
J. Phys. B {\bf 35}, 1115 (2002).

\bibitem{And02b},
K.~Andrae, A.~Pohl, P.-G.~Reinhard, C.~Legrand, M.~Ma, and E.~Suraud.
in {\em {Progress in Nonequilibrium Green's Functions {II}}}, p.28,
Edts M.~Bonitz and D.~Semkat, World Scientific, Singapore, 2003.

\bibitem{Ger04a}
B.~Gervais, E.~Giglio, P.-G. Reinhard, and E.~Suraud.
\newblock {Phys. Rev. A}, {\bf 71}, 015201 (2004).

\bibitem{Gig01a}
E.~Giglio, P.-G.~Reinhard, and E.~Suraud.
\newblock{J.~Phys.~B} {\bf 34}, 1253 (2001).


\end{thebibliography}

\end{document}